# Dynamic Red Queen explains patterns in fatal insurgent attacks

Neil Johnson[1], Spencer Carran[2,3], Joel Botner[4], Kyle Fontaine[5], Nathan Laxague[1], Philip Nuetzel[5], Jessica Turnley[6] and Brian Tivnan[7,8]

[1] *Physics Department, University of Miami, Coral Gables, Florida 33124, U.S.A.*
[2] *Biology Department, University of Miami, Coral Gables, Florida 33124, U.S.A.*
[3] *Mathematics Department, University of Miami, Coral Gables, Florida 33124, U.S.A.*
[4] *Computer Science Department, University of Miami, Coral Gables, Florida 33124, U.S.A.*
[5] *International Studies Department, University of Miami, Coral Gables, Florida 33124, U.S.A.*
[6] *Galisteo Consulting Group, Albuquerque, New Mexico 87110, U.S.A.*
[7] *The MITRE Corporation, McLean, VA 22102, U.S.A.*
[8] *Complex Systems Center, University of Vermont, Burlington, VT 05405, U.S.A.*

## ABSTRACT

**The Red Queen's notion "It takes all the running you can do, to keep in the same place" has been applied within evolutionary biology, politics and economics. We find that a generalized version in which an adaptive Red Queen (e.g. insurgency) sporadically edges ahead of a Blue King (e.g. military), explains the progress curves for fatal insurgent attacks against the coalition military within individual provinces in Afghanistan and Iraq. Remarkably regular mathematical relations emerge which suggest a prediction $\tau_n = \tau_1 n^{-[m\log_{10}\tau_1 + c]}$ for the timing of the *n*'th future fatal day, and provide a common framework for understanding how insurgents fight in different regions. Our findings are consistent with a Darwinian selection hypothesis which favors a weak species which can adapt rapidly, and establish an unexpected conceptual connection to Physics through correlated walks.**

How has the Afghanistan insurgency consistently managed to increase the frequency of fatal attacks against a far stronger force – the coalition military? Inspired by recent suggestions from the evolutionary biology community that fast adaptation is a key to survival (*1-3*), we use progress curves (*4-15*) to analyze the daily 'production' of military fatalities by the insurgency within individual provinces in Afghanistan and Iraq. Progress curves (or 'functions') (*5*) are a widely used tool for quantifying how human productivity changes over time, e.g. in industrial manufacturing (*4-7*), software installation (*10*) and even cancer surgery (*11-15*). The province-by-province progress curves reveal a remarkable regularity that we explain in terms of a dynamical arms race between an adaptive Red Queen (i.e. insurgents) and her counter-adapting Blue King opponent (i.e. coalition military). Practical consequences include a tool that could help predict future fatal days for the military, and a new perspective on how insurgent conflicts evolve at the level of individual provinces. More broadly, our work identifies potentially generic dynamics for any competitive interaction between asymmetric populations (e.g. virus versus immune system), and establishes new connections between the search for effective counterinsurgency strategies, evolutionary biology, industrial manufacturing, and the physics of correlated walks (*16-27*).

Our data and analysis are freely available for public scrutiny: The coalition military fatality data come from the publicly accessible website www.iCasualties.org, which is highly regarded as reliable and independent (*28*), while our analysis is performed using the free downloadable tool Open Office which runs on any computer platform (see Appendix). For Afghanistan, we include fatalities from the start of Operation Enduring Freedom in 2001 until the breakpoint in summer 2010 when Gen. Petraeus became ISAF Commander and the U.S.-led surge started. For Operation Iraqi Freedom, the data includes fatalities from 2003 until summer 2010 when U.S. military action officially ended.

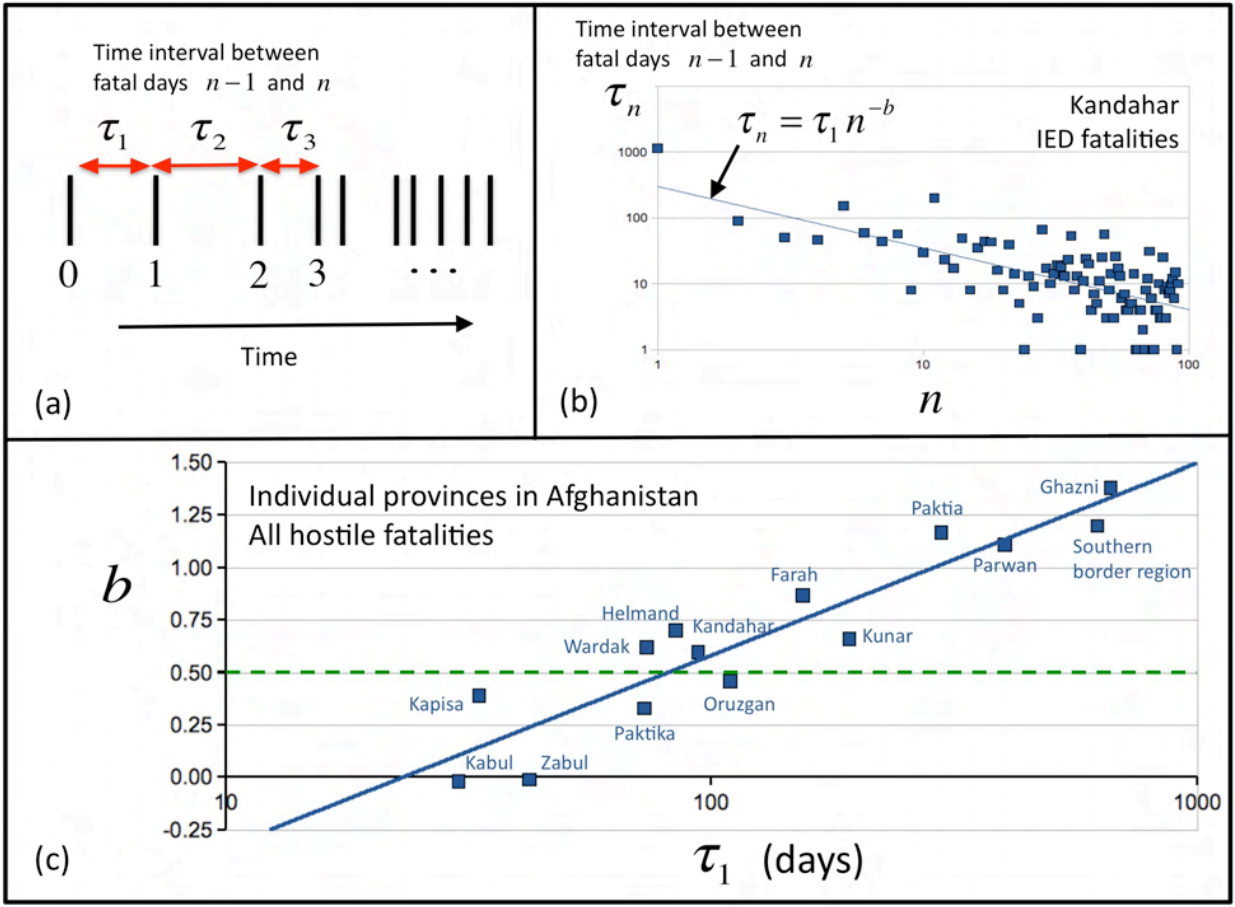

Figure 1: Insurgency progress curves. (a) Schematic timeline of successive fatal days (shown as vertical bars) for the coalition military. $\tau_1$ is the time interval between the first two fatal days, labelled 0 and 1. (b) Example of a best-fit progress curve, for the individual province Kandahar and with data filtered for military IED fatalities only. On the log-log plot, the best-fit progress curve is a straight line with slope $-b$ and intercept $\tau_1$. Empirical datapoints correspond to successive time intervals $\tau_n$ between fatal days. (c) Plot shows $b$ and $\tau_1$ for individual provinces including all hostile fatalities (i.e. all military fatalities attributed to insurgent activity). The simple coin-toss limit of our dynamic Red Queen theory predicts $b = 0.5$ exactly (i.e. green dashed line). More generally, our theory predicts $b \geq 0$ with $b$ values spread around 0.5, in agreement with what is observed here. The subset of fatalities recorded in iCasualties as 'Southern Afghanistan' are shown as a separate region because of their likely connection to operations near the Pakistan border.

Figures 1(a) and (b) show how we use the standard progress curve form $\tau_n = \tau_1 n^{-b}$ (4,5) to describe the non-stationary nature of fatal days for the coalition military within each province. We define a fatal day for a particular province P as one on which at least one coalition military died as a result of hostile activities. Progress curves are obtained for fatal days resulting from any hostile activity, and for particular subsets filtered by method (e.g. IED). For the scenario that the data follows the progress curve perfectly in Fig. 1(b), $\tau_1$ is the time interval between the first two fatal days labelled $n=0$ and $n=1$. *The value of $b$ is of crucial practical importance since it determines how fast the time interval between consecutive fatal days decreases* (see Fig. 1(b)). The insurgents would like $b$ to be large and positive so that subsequent time-intervals $\tau_n$ decrease quickly from the initial value $\tau_1$, while the military wish $b$ to be negative so that $\tau_n$ actually increases. The amount of scatter in Fig. 1(b) is consistent with progress curves from many non-conflict applications including successive operations by a medical surgery team (*10-15*), and is to be expected given the struggle underway. The minimal timescale is one day (i.e. $\tau_n \geq 1$) since iCasualties does not contain intraday detail. We do not consider non-fatal events (i.e. no coalition military casualties) since they occur almost every day and so do not generate interesting progress curves (i.e. $\tau_n \approx 1$ for all $n$). Alternative progress-curve analytic forms have been debated *(8)* -- however any such two-parameter progress curve amounts to a complex nonlinear transformation of the conventional power form that we use, and hence simply generates a more complex version of Fig. 1(c). We have checked alternative forms, but we find their fits to be no better than the conventional power form.

Figure 1(c) presents the remarkable linear relationship that we uncover for the $b$ and $\tau_1$ values of different provinces. The best-fit straight line in Fig. 1(c) is $b = m \log_{10} \tau_1 + c$ where $m = 0.91$, $c = -1.25$ with a high value of the Pearson rank product-moment correlation coefficient $R^2 = 0.9$. This linear pattern is unexpected given that individual provinces differ dramatically in terms of (i) the numbers of fatalities, with a province such as Kandahar in the south having many more fatalities than Farah in the west; (ii) the calendar dates for their respective initial military fatality ($T_{n=0}$) and subsequent time intervals $\tau_n$, e.g. the first recorded military fatality in Paktia was on 4 March 2002 while in Wardak it was 25 July 2007; (iii) their distinct individual geographies, tribal ethnicities and histories *(29)*; (iv) the numbers, nationalities and distributions of coalition troops within each province. We conclude (see Appendix) that this pattern is unlikely to be driven by international or national events, including troop increases, since the progress curve for the timelines of such events is a very poor fit, while even the best-fit $(b,\tau_1)$ values differ substantially from those in Fig. 1(c). In particular, it would require a very specific province-dependent surge in troops and/or insurgents *and* a very contrived subsequent mathematical relationship to military fatalities, to reproduce these provincial interdependencies. Going further, if fatal days in different provinces were synchronized in calendar time, they should all sit at the *same* point – but they do not. If instead, fatal days in different provinces were independent, then the individual $(b,\tau_1)$ points should be scattered *randomly* across Fig. 1(c) – but again, they are not. It is also curious that only about one half of all provinces have enough fatal events to produce a meaningful progress curve and hence appear in Fig. 1(c).

Figure 1(c) has an immediate consequence for predicting fatal days for the military. Suppose a previously peaceful province P has its first military fatalities on days $T_{n=0}$ and $T_1$, but the subsequent evolution is unknown. We start by obtaining the best linear fit in Fig. 1(c) without province P, hence obtaining estimates for $m$ and $c$. We then insert $\tau_1$ (i.e. $\tau_1 = T_1 - T_{n=0}$) into $b = m \log_{10} \tau_1 + c$ to obtain an estimate of $b$ for province P. We then generate the time interval between the $(n-1)$'th and $n$'th future fatal days in P using the progress curve formula $\tau_n = \tau_1 n^{-b}$. In other words, by substituting the formula $b = m \log_{10} \tau_1 + c$ obtained from Fig. 1(c) without P into the progress curve $\tau_n = \tau_1 n^{-b}$, we obtain the prediction formula $\tau_n = \tau_1 n^{-[m \log_{10} \tau_1 + c]}$ for province P in which the only input is the time interval between the first two events $\tau_1$. Adding up prior $\tau_n$ values yields the predicted time-interval between the first fatal day in P (i.e. $T_{n=0}$) and the $n$'th. We stress that this $n$'th fatal day might lie months or years in the future.

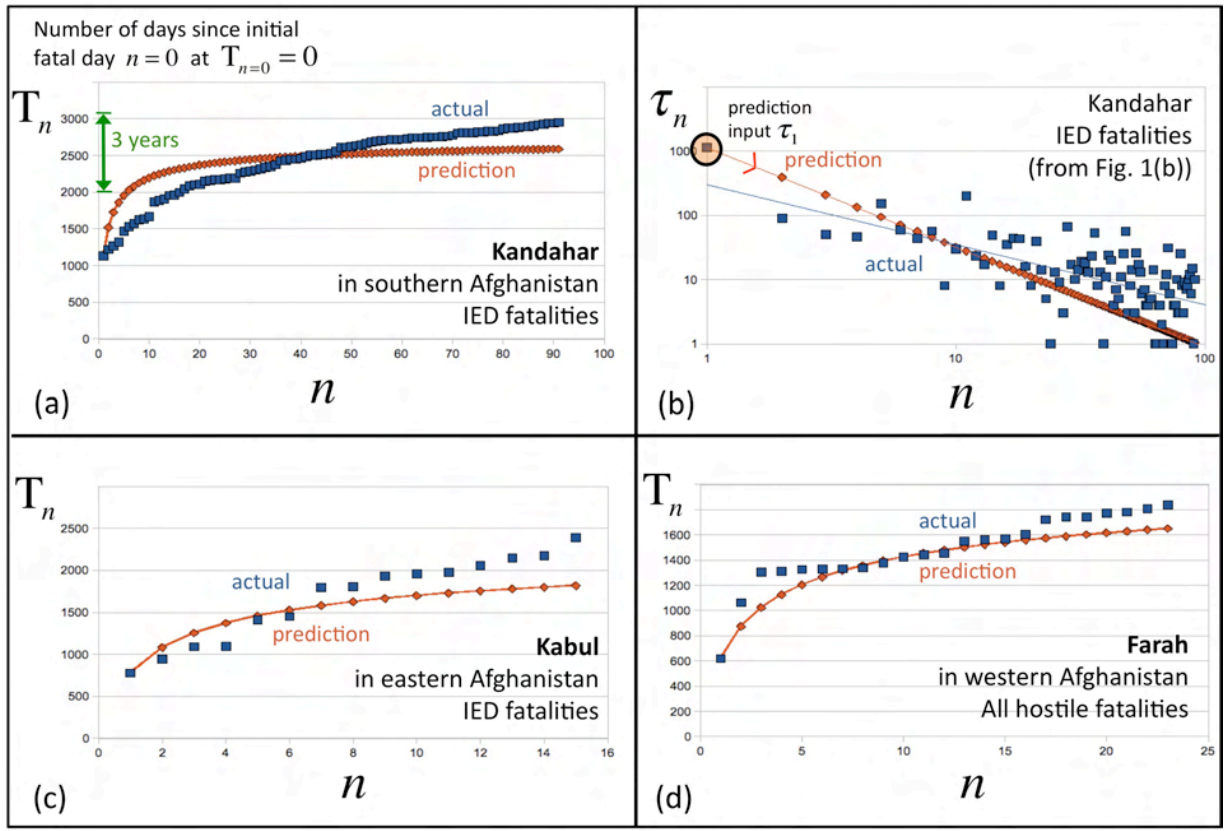

Figure 2: Predicting future fatal days for the coalition military. (a) The predicted (red diamonds) vs. actual (blue squares) calendar time $T_n$ in days, between the initial day ($T_{n=0} = 0$) on which there was a military IED fatality in Kandahar, and the $n$'th subsequent day having a military IED fatality. The only information from the iCasualties Kandahar data which we use as input for the prediction, is the first empirical time interval $\tau_1$. An interval of 3 years is shown on the $T_n$ axis to highlight the long-term nature of the prediction. Red line is a guide to the eye. (b) Comparison of predicted time intervals to the actual ones. (c) Same as (a) but for Kabul. (d) Same as (a) but for Farah, and now including all hostile military fatalities (i.e. fatalities attributed to insurgent activity).

Figure 2 illustrates our prediction results for three provinces which are widely separated geographically, from western (Farah), southern (Kandahar) and eastern (Kabul) Afghanistan. Figures 2(a)-(c) illustrate our results for IED fatalities while Fig. 2(d) considers all hostile fatalities. The agreement is surprisingly good, given that we are predicting fatal *days* which may be several *years* into the future, and that the only input is the time interval between the first two events $\tau_1$ which will understandably be noisy for an insurgent conflict. Indeed, the prediction accuracy for IED fatalities is achieved by estimating $b$ from the all-hostile curve without having to filter out non-IED events and hence suffer from sparser data, i.e. we simply used the all hostile results in Fig. 1(c) without province P. Other Afghanistan provinces show fairly similar results, as does pre-surge Iraq discussed later, with even the worst-case prediction having the correct order of magnitude. Our prediction assumes that the progress curve does not experience any significant structural changes, hence the fact that the empirical data tends to move above the predicted curve as $n$ increases, suggests that the military's ability to disrupt insurgency activity is increasing over time. In addition to operational planning, this prediction tool can be used to estimate the likely trajectory of fatal attacks in a region which has, so far, been relatively quiet (e.g. only two fatal days so far, and hence only $\tau_1$ is known). If a significant deviation suddenly arises between the predicted and actual curve, this serves as an indicator that the conflict in that region is experiencing a fundamental shift (see Appendix). We have found that the subsequent prediction accuracy can then be restored by restarting the count of fatal days from that later time. Depending on the availability of reliable daily data, future work will test this scheme on other conflicts.

One might argue that any such prediction scheme is useless once known publicly since it can be invalidated by insurgents' free will. However we believe that this will not happen for the same reason that all commuters know that a traffic jam will appear every day at rush hour on a certain route, yet many still end up joining it. External constraints of working hours, school schedules and finite numbers of direct roads, mean that such predictability is hard to avoid. Similarly, the spontaneity of fatal attacks by the insurgency is likely constrained by many factors, including the availability of troop convoys, explosive materials, and sympathy within the local population.

Figure 3 shows that the quasi-linear relationship that we found in Fig. 1(c) is surprisingly generic. Specifically, it shows the more specific case of IED fatalities and includes provinces from both Afghanistan and Iraq. It is understandable why the scatter is larger than in Fig. 1(c) given the inclusion of two separate countries, and the sparser number of IED fatalities as compared to total hostile fatalities. Indeed, some provinces have insufficient IED events to appear. The fact that there is no obvious connection between the Iraq and Afghanistan conflicts other than a common coalition military (i.e. Blue King), makes the linear trend even more of a puzzle and heightens the need for an explanatory quantitative theory. The recent unveiling of Wikileaks (*30*) gives us an independent check of our iCasualties data at the country level, with Fig. 3 demonstrating that the $b$ and $\tau_1$ values are essentially the same for both sources. Within individual Iraqi provinces, the progress curves exhibit a shift point $n_{shift}$ separating their initial behavior which is typical of the Afghanistan conflict (i.e. Fig. 1(b)), from a regime toward the end of the Iraq conflict where $\tau_n$ generally increases. Crudely speaking, this shift occurs around the time of the military surge. In order to compare like with like, we only include pre-shift Iraq results (i.e. $n < n_{shift}$) in Fig. 3 and defer brief comments on post-shift Iraq to the Appendix. We determined $n_{shift}$ for each province by calculating the turning point in the cumulative moving-average of the time interval $\tau_n$, but we have checked that our results are insensitive to the precise value of $n_{shift}$.

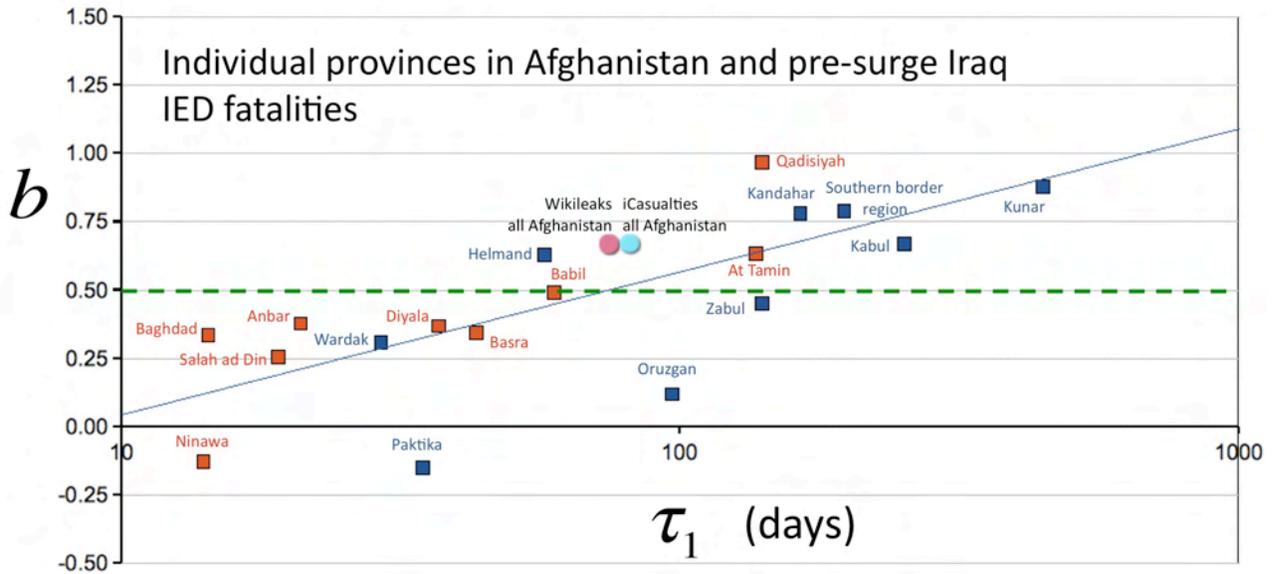

Figure 3: Progress curve parameter values for distinct provinces in Afghanistan (blue) and pre-surge Iraq (red) for IED fatalities. Results show a similar quasi-linear dependence to Fig. 1(c), despite the inclusion of provinces from two separate conflicts in two separate countries. Since the frequency of IED fatalities for the provinces sitting just below $b=0$ is so low, their $b$ values are very uncertain and hence should be regarded simply as $b \approx 0$. As in Fig. 1(c), the simple coin-toss limit of our dynamic Red Queen theory predicts $b=0.5$ exactly (i.e. green dashed line). More generally, our theory predicts $b \geq 0$ with $b$ values spread around $0.5$, in agreement with what is observed here. Pink circle shows Wikileaks value for IED fatal days aggregated over all Afghanistan, while blue is for iCasualties.

Identifying the precise mix of province-dependent socioeconomic, tactical and operational factors that dictate the details in Figs. 1(c) and 3, throws up a fascinating challenge for the socioeconomic and military analysis community. We refer to Ref. 1 for a detailed discussion of some likely issues -- however we now show that the concept of a dynamical arms race between adaptive insurgents (Red Queen) and the coalition military (Blue King) offers a quantitative answer to two key questions: (1) Why does the pattern of fatal days for the military appear to follow a progress curve $\tau_n = \tau_1 n^{-b}$? (2) Why do the $b$ values have a broad scatter around 0.5, despite the fact that *any $b$ value from zero to infinity* could in principle have emerged?

We start by defining $R$ to be the Red Queen's lead over her opponent (Blue King). $R$ represents a real distance in a running race, or a more abstract strategic advantage in an arms race. The traditional Red Queen story involves her running as fast as she can in order to stay at the same place *(31,32)*. This implies that the Blue King instantaneously and perfectly counter-adapts to any Red Queen advance, such that they are always neck and neck (i.e. $R=0$). However, such instantaneous and perfect counter-adaptation would be unlikely in practice. Reference 1 suggests that the Red Queen insurgency may adapt faster than her larger, and possibly more sluggish, Blue King military opponent. The Red Queen could then sporadically edge ahead (i.e. increase $R$) as a result of fast adaptation to new technology, information, past experience etc., and/or the inability of the Blue King to instantaneously reduce $R$ again. The Red Queen's lead $R$ would then undergo a jerky but gradual increase. Even if the actual dynamics of $R$ were known, they would likely be so complex as to appear seemingly random. This motivates us to exploit a trick commonly used in Physics *(27)*: We simply mimic this complex, jerky 'walk' using repeated tosses of a coin, where throwing a heads increases $R$ by one while a tails reduces $R$ by one. We hence generate an $R$ 'walk', e.g. $R = \ldots,3,4,5,4,5,6\ldots$ (see Appendix). Assuming $R$ starts off somewhere near zero, tossing a coin will produce walks where roughly half have $R$ predominantly positive (and hence the Red Queen is predominantly in the lead) while the other half have $R$ as predominantly negative (and hence the Blue King is predominantly in the lead). The negative walks represent provinces in which the Blue King is predominantly ahead of the Red Queen and hence the number of fatal days is negligibly small, thereby explaining why half or more provinces in Afghanistan and Iraq have *insufficient* fatal days to appear in Figs. 1(c) and 3 (e.g. only 9 out of 18 Iraq provinces listed). Since we wish to explain the behavior of Figs. 1(c) and 3, we henceforth focus on the case where $R$ is predominantly positive, i.e. the Red Queen sporadically edges ahead of the Blue King.

The larger $R$ becomes, the larger the insurgent advantage and hence the more likely that any particular day has military fatalities. We therefore assume that the size of $R$ is a proxy for the insurgents' rate of production of fatal days at that moment. It is also reasonable to expect that large increases or decreases in $R$ will be focused around the time of fatal days. Insurgents have by definition become successful at that moment and this may help generate a further increase in $R$, while the military's loss of life may force it to step up counter-adaptation efforts to seek an imminent decrease in $R$, hence $R$ is predominantly a function of $n$ (i.e. $R(n)$). An exact mathematical result from Physics *(27)* is that the typical size of $R$ after $n$ steps is given by $n^b$, with $b = 0.5$ for a walk generated by coin tosses (see Appendix). For *correlated* coin tosses, a walk emerges with $n^b$ but where the $b$ value depends on the correlations introduced *(27)*. The fact that $b = 0.5$ acts as a crude midpoint value for all such cases, can be understood by considering two extreme limits: If military counteradaptation is completely inadequate or absent, the insurgents' advantage $R(n)$ will increase at every step $n$ and hence $R(n)$ varies as $n^b$ with $b = 1$, i.e. the Red Queen now walks uninhibited like somebody walking down a street, hence her lead $R(n)$ is proportional to $n$ which is mathematically equivalent to $n^b$ with $b = 1$. Indeed, if she picks up momentum, she may even start accelerating and hence $b > 1$ as seen in Figs. 1(c) and 3 for a few provinces. By contrast, effective military counteradaptation to each Red Queen advance means that $R(n)$ stays close to zero, hence $R(n)$ varies as $n^b$ with $b \approx 0$. We stress that it is only in the idealized case where the Blue King's counter-adaptation is instantaneous and perfect, that $R(n)$ is always exactly zero; and it is only if the Blue King proactively produces its own advances, that $R(n)$ becomes negative and hence the province becomes peaceful (which is what happens for post-surge Iraq, see Appendix).

Since $R(n)$ acts like the insurgents' rate of production of fatal days, and the rate of fatal days is inversely proportional to the time interval $\tau_n$, we have shown that $\tau_n$ varies as $n^{-b}$ with $b \geq 0$. We have also shown that although $b \geq 1$ is possible, the value $b = 0.5$ corresponding to the coin-toss walk represents a crude midpoint (green dashed line in Figs. 1(c) and 3). Determining the constant of proportionality by setting $n = 1$ in $\tau_n \propto n^{-b}$, we obtain $\tau_n = \tau_1 n^{-b}$. Our dynamic Red

Queen theory has therefore answered questions (1) and (2), and hence provides a quantitative explanation for the main features of Figs. 1(c) and 3. Its success also opens the door to addressing more general operational issues (e.g. timing interventions) by exploiting known results from the physics of biased-walks *(27)*. (See Appendix for an example of a multi-faceted Red Queen advantage).

To explore why individual provinces appear in the order that they do in Figs. 1(c) and 3, we carried out a comparison to freely available sociotechnical maps *(29)*. The lack of correlation that we found with obvious candidates such as geographical proximity, proximity to Pakistan, high density of internally displaced persons, common tribal warlords and significant poppy production, suggests that each can be ruled out as the dominant determinant. For example, although geographical neighbors Kandahar and Helmand are next to each other in Fig. 1(c), the geographical neighbors Parwan, Wardak and Kabul are widely separated. By contrast, Farah and Kunar are next to each other in Fig. 1(c) but lie on opposite sides of Afghanistan, hence ruling out any simple contagion mechanism. We do however believe that the appearance of linear trends in Figs. 1(c) and 3 sheds light on the *way* in which the insurgents fight from province to province, and country to country – in particular, it reveals a loose operational coupling between provinces within the same country, and between countries, which could result from the fact that the insurgents within each province are fighting a common Blue King opponent and hence copy successful tactics from elsewhere, as suggested by several recent narratives (see *(1,25)* and Appendix). When MRAPs (Mine Resistant Ambush Protected vehicles) were moved to Afghanistan, for example, the Blue King effectively took on very similar operational characteristics in each country, i.e. slow, lumbering movement restricted to certain transport corridors. Further support comes from a numerical simulation we have performed in which we couple the $R(n)$ values for a set of Red Queen-Blue King races, to mimic different provinces. As in Figs. 1(c) and 3, a province starting off with infrequent fatal days tends to become violent quicker (i.e. high $\tau_1$ leads to high $b$) while a province starting off with frequent fatal days evolves relatively slowly (i.e. low $\tau_1$ leads to low $b$). This dynamical coupling suggests fruitful future links to the topics of firefly entrainment (*33,34*), coupled ecological patch models *(35),* non-Poissonian two-way human correspondence *(36)* and team collaborations *(37)*.

Our dynamical Red Queen-Blue King mechanism should be applicable to other two-population interactions with asymmetries in power and adaptive speed (e.g. virus-immune system) *(1,32)*. It could also apply to the original progress-curve domain of manufacturing for scenarios where the task's difficulty changes through feedback, e.g. the Red Queen plays the role of a manufacturer while the Blue King represents some production machinery or process with time-varying properties *(11-15)*.

**Acknowledgements**: The authors gratefully acknowledge support for this research from the Joint IED Defeat Organization, IDN# N70465, and from The MITRE Corporation and the Santa Fe Institute for their co-hosting of the "Mathematics of Terrorism" workshop. The views and conclusions contained in this paper are those of the authors and should not be interpreted as representing the official policies, either expressed or implied, of any of the above named organizations, to include the U.S. government. We are also grateful to P. Dodds, C. Danforth and A. Clauset for broad discussions surrounding this topic and to L. Amaral for earlier discussions concerning non-Poissonian behavior.

# Appendix

## 1: Steps to reconstruct and verify our results

Here we describe how anyone with Internet access can process the freely available casualty data at www.iCasualties.org in the way we did, and analyze these data using the free software Open Office software which runs on any platform. We start with a note about our data. We downloaded our dataset from iCasualties in the summer of 2010. We downloaded the full raw dataset, and applied our own filters from within our spreadsheets. As with all active datasets, it is possible that the dataset that we downloaded is subsequently changed slightly and/or updated by the database manager – for example, new drop-down filters have now been added -- but likewise, it is also possible that such additional handling may introduce unwanted errors and that the filters are incomplete. For this reason, we decided to work with the full raw data as it stood in summer 2010, and introduce our own user-defined filters that we could double-check ourselves. For this reason, results from later dataset versions and/or using the online filters may not be exactly the same as ours, though we expect any differences to be minor. In any case, our dataset is available to any reader who wishes to access it.

A. DATA PROCESSING

1. Afghanistan:

All data were prepared in OpenOffice.org Calc, starting with the "All_Data_Afghanistan" sheet from "iCasualties_Raw_Data_v2.ods."

For the file "Afghan_Filtered.ods," the following filters were used:

All IED: Standard filter, reason contains " ied" (with a space)

Hostile && !FF: Standard filter, reason does not contain "non-hostile" AND reason does not contain "non hostile" AND reason does not contain "friendly"

Filters for Provinces are self-explanatory.

Repeat days in the same province are filtered by setting the filter "tau>0." The time lag is computed by setting the formula '=a3-a2' and dragging it down for all other rows. We removed the two soldiers in Afghanistan who died of "unknown" causes from the filtered data, and in All Hostile we exclude friendly-fire deaths since these are accidents. Although slightly different filters can be applied to capture All Hostile, we have checked that our main conclusions do not change significantly according to filter.

The document "Afghan_Filtered_Tau.ods" is the same as "Afghan_Filtered.ods" except that, as implied in the name, the former is filtered for tau>0.

2. Iraq:

All data were prepared in OpenOffice.org Calc, starting with the "All_Data_Iraq" sheet from "iCasualties_Raw_Data_v2.ods."

For the file "Iraq_Filtered.ods" the following filters were used.

All IED: Standard filter, reason contains " ied" (with a space)

All Hostile: Standard filter, reason does not contain "non-hostile" AND reason does not contain "Not reported yet" AND reason does not contain "friendly"

Filters for Provinces are self-explanatory. The time lag is computed by setting the formula '=a3-a2' and dragging it down for all other rows. Multiple events on the same date are combined by filtering for tau>0 and copying to a new sheet (to correct row numbers).

The document "Iraq_Filtered_Tau.ods" is the same as "Iraq_Filtered.ods" except that the former is filtered for tau>0.

B. OPEN OFFICE ANALYSIS

From the files "Iraq_Filtered_Tau.ods" and "Afghan_Filtered_Tau.ods" the following procedure is used to obtain development curves within OpenOffice.org Calc:

On the sheet containing the data of interest (depending on what province and what cause of death) highlight all the tau values, place on a log-log scale, and create a power fit. See main paper for a discussion of the shift between pre and post-surge behaviors.

In Iraq, it is necessary to first define an n-shift date before making development curves. The n-shift is defined as the minimum value of the cumulative moving average after the first 15 events, except in Qadisiyah which only has a total of 15 IED events, thus the global minimum is used there. The file "pre_nshift_devcurves.ods" contains these development curves, with the $n_{shift}$ dates for each province highlighted.

Once power regressions of the form $A x^{-b}$ are made, a plot of log(A) vs b can be generated. For a given province, generate a log(A) vs b plot that omits its data, and obtain a linear regression b=mlog(A)+c. Input the provinces first tau value for A and thus obtain a prediction for b. This can then be passed into the prediction formula as described in the main paper.

Wikileaks Data:
The file "Wikileaks_Afghan.ods" contains the entire IED-related database released by Wikileaks. From this file, filter for "FriendlyKIA>0" and then generate tau values as before. Filter for tau>0 and generate a power regression. This should be equivalent to the development curve from IED events over all provinces in Afghanistan, and provides a check for how much the Wikileaks and iCasualties databases differ. As confirmed in Fig. 3, essentially the same results are obtained from both for the progress curves aggregated across all provinces. Since the province information in Wikileaks tends to be given in terms of coordinates as opposed to names, we could only compare to the iCasualties data at the level of the entire country. The fact that the two sources are then in excellent agreement is reassuring, given the possible problems with any casualty dataset gathered under conflict conditions.

## 2: Analysis of timelines of events etc.

The timeline events are from
http://www.historycommons.org/timeline.jsp?timeline=complete_911_timeline&startpos=0&complete_911_timeline__war_on_terrorism__outside_iraq=afghanistan ,
http://www.securitycouncilreport.org/site/c.glKWLeMTIsG/b.2687219/k.5FAA/Afghanistanbr_Historical_Chronology.htm, http://news.bbc.co.uk/2/hi/1162108.stm,
as well as the series on Wikipedia starting with 2001
http://en.wikipedia.org/wiki/2001_in_Afghanistan
On the bottom of this page it has links for 2001, 2002, 2003, etc. up to 2010. We picked major categories: Major Operations/Battles, Change In Leadership (assassinations or change in UN commanders), Change In Policy (in terms of the UN forces, the USA, or even the Afghan government), Change In Civil Society (things that would either make life better for civilians and therefore less likely to be insurgents or the opposite effect, such as Internet connection, or the ending of the curfew), Elections, Increase In Insurgent Pool (events like the prison break that could increase the insurgent forces), Natural Disasters, and Troop Surges. After picking these categories, we methodically went through every single event on the aforementioned timelines and picked out the ones we judged as most likely to be relevant. In this way, we tried to minimize noise. The figure below shows the results (i.e. orange bars) for all these categories combined – however, similar conclusions are obtained for just individual (or combinations of subsets of) categories. We could find no combination of these exogenous timelines that could be seen as a nucleation of the military fatality timelines. In particular, when we calculated the best-fit progress curves, the fits were not only very poor (far poorer than the typical fatal day data), but they also gave rise to $(b, \tau_1)$ values that were qualitatively different from the ones shown for individual provinces in Figs. 1(c) and 3 – both for all hostile fatalities and also just IED fatalities. Hence the comment in the main paper that these events do not seem to be the dominant driving force. Instead, their effect presumably feeds in with all the other factors in some complex way that remains to be determined.

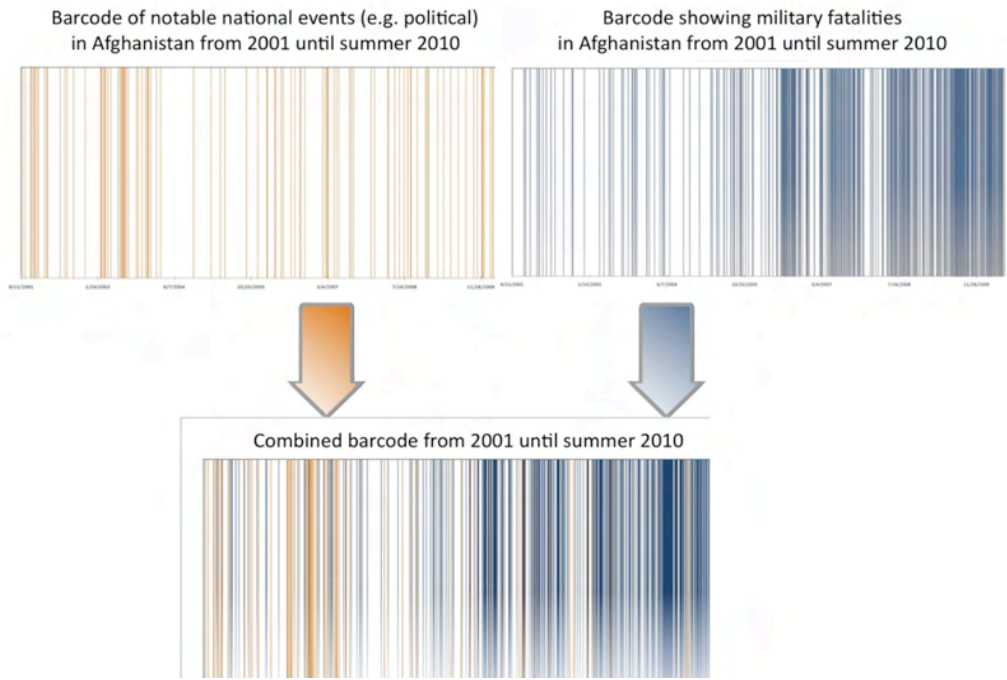

Figure caption: Timelines for Afghanistan. Orange corresponds to relevant national events (e.g. political) extracted according to recipe in Appendix. Blue corresponds to fatal days from insurgent attacks. There is no strong evidence to suggest that these exogenous events (orange) nucleate the military fatal days (blue). In particular, not only is the progress curve a poor fit to the national events (worse than the fit to fatal days) but also the best-fit parameter values $b$ and $\tau_1$ are very different from the ones shown for fatal days in Figs. 1(c) and 3.

## 3: Narrative concerning loose coordination

In addition to the feature of a common enemy (i.e. coalition military), we give here some example narratives which support the idea of an indirect, loose coordination between Afghanistan and Iraq, and hence which offer an explanation for the prominent quasi-linear pattern describing the combined Afghanistan-Iraq datapoints in Fig. 3:

Insurgents going from Afghanistan to Iraq:
http://en.wikipedia.org/wiki/Al-Qaeda_in_Iraq

Insurgents going from Iraq to Afghanistan:
http://www.nytimes.com/2008/10/15/world/asia/15afghanistan.html
http://www.independent.co.uk/news/world/asia/remember-afghanistan-insurgents-bring-suicide-terror-to-country-523323.html

## 4: Post-surge Iraq
As noted in the main paper, the progress curves within individual Iraqi provinces exhibit a shift point $n_{\text{shift}}$ separating two regimes. Their initial behavior is typical of the Afghanistan conflict (i.e. Fig. 1(b)), while a second regime arises toward the end of the Iraq conflict and features $\tau_n$ generally increasing. In short, the fatal days in post-surge Iraq exhibit a general increase in $\tau_n$. The overall behavior tends to mirror the pre-surge results but with *negative* $b$ values if we reset the fatal day count (i.e. set $n=0$) to start from $n_{\text{shift}}$. In addition to any intrinsic operational interest, this concurrence of an increase in $\tau_n$ with an increase in troops demonstrates that more troops does not necessarily equate to more fatal attacks (and hence more frequent fatal days). In other words, *the patterns we find in the progress curve behavior are not simply linked to an increase in the number of troops and hence an increase in the number of targets.*

We have tried many different ways of defining the shift value $n_{\text{shift}}$, along the theme of moving averages with different window sizes etc. We found that neither the precise choice – nor the precise definition -- have much affect on the progress curve results show in Fig. 3. We chose a cumulative moving average for Fig. 3 since it is a common tool in spreadsheet programs and has a concrete implementation, hence can be checked easily by readers.

# 5: Mathematical details underlying Red Queen walk $R(n)$

## 5.1 A coin-toss walk

Suppose we toss a fair (i.e. unbiased) coin. A coin has no memory of past outcomes, which is another way of saying it has no *correlations* in the time-series of its outcomes. Heads (H) and tails (T) have equal probability of occurring, and this is independent of what outcomes have previously arisen. We denote the coin-toss outcome at a given step $i$ as having value $\Delta x_i$, where heads generates $+d$ while tails generates $-d$. The successive outcomes $\{\Delta x_i\}$ are i.i.d. variables, which means independent and identically distributed. An example of such a coin-toss 'walk' is shown below for a series of outcomes HHTH with step-size $d$. The walk position corresponds to the value of the Red Queen lead's $R(n)$, and is shown as a thick black line together with the 'tree' formed by all possible such walks:

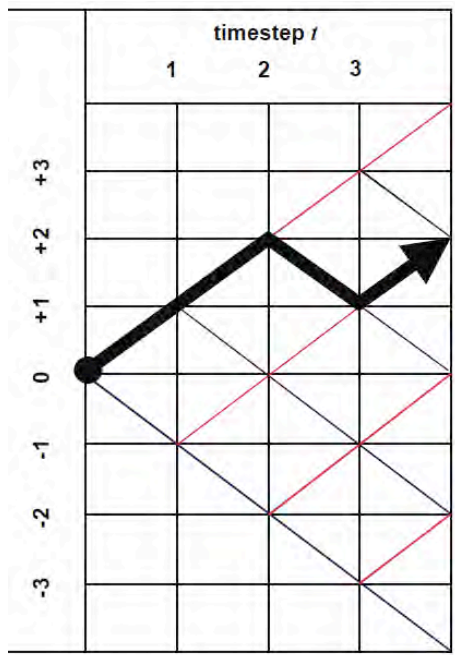

Figure caption: Red Queen's walk $R(n)$ in units of $d$, as a function of the step number $n$ (i.e. each step corresponds to $n \to n+1$). In the simple scenario, her lead is generated by a coin-toss where an increase or decrease of magnitude $d$ occurs with probability 0.5 at each step. Thick black line corresponds to the sequence of coin-tosses HHTH. All other possible paths form a 'tree' as shown. As explained in the main paper, peaceful provinces correspond to $R(n)$ predominantly below zero (i.e. $R(n) < 0$ since the walk is then dominated by Blue King advances) while violent ones have $R(n)$ predominantly above zero (i.e. $R(n) > 0$ since the walk is then dominated by Red Queen advances).

As discussed in the main paper, the provinces appearing in Figs. 1(c) and 3 are by definition ones with fatal days. Since we wish to explain their features, we will focus on walks with $R > 0$ since these are the ones dominated by Red Queen advances and hence are the ones which are likely to produce enough fatal days to appear in Figs. 1(c) and 3. Whether the walk ends up with $R > 0$ by insurgents' good fortune (i.e. a winning streak of Heads being tossed), or whether it is because of clever tactics, does not affect the mathematics. Because of the unbiased nature of a coin-toss, approximately half of all walks will end up in the region $R > 0$ (i.e. walk is dominated by Red Queen advances). This suggests that approximately half of all provinces should have enough fatal days to show up in Figs. 1(c) and 3. This is indeed roughly what is observed, as mentioned in the main paper. The other half of all walks end up in the region $R < 0$. These are the ones where the Blue King dominates, and these mimic a non-violent province. The traditional situation of the Blue King *instantaneously* and *perfectly* counteradapting to every Red Queen move, just as she is about to make it, is impossible in practice -- and also does not arise in this model, since we have successive coin-toss outcomes and *somebody* has to win at each step. Even if the Blue King always catches up in the next step, this is not *instantaneous* adaptation-counteradaptation. A 'timestep' in the coin-toss model is, as explained in the main paper, a step from a given $n$ to the next value $n+1$ since fatal days are assumed to act as the dominant ticks of the clock.

## 5.2 Statistical properties of $R(n)$

We now consider the statistical properties of the average and variance for $R(n)$ after $n$ steps. Our notation for the change in $R(n)$ at step $i$ is $\Delta x_i = x_i - x_{i-1}$. Hence the change between step 0 and $n$ is given by $\Delta x_{n,0} = \sum_{j=1}^{n} \Delta x_j = x_n - x_0$. The *mean* change between step 0 and $n$ is:

$$\langle \Delta x_{n,0} \rangle = \sum_{j=1}^{n} \langle \Delta x_j \rangle \tag{5.1}$$

which is the well-known result that the *average of the sum is equal to the sum of the averages*. Equation (5.1) holds *irrespective* of whether the changes $\Delta x_j$ are i.i.d. or not. For the special case in which each mean is always the same $\langle \Delta x_j \rangle \equiv \langle \Delta x \rangle$ (for example, for i.i.d. variables) then we have:

$$\langle \Delta x_{n,0} \rangle = \sum_{j=1}^{n} \langle \Delta x_j \rangle = n \langle \Delta x \rangle \tag{5.2}$$

For the case of the coin-toss walk above, we have $\langle \Delta x \rangle = 0$ where the average includes all possible walks in the 'tree' (see figure). Hence $\langle \Delta x_{n,0} \rangle = 0$ for all $n$. The *variance* is as follows:

$$\sigma_{n,0}^2 \equiv \left\langle \left(\Delta x_{n,0} - \langle \Delta x_{n,0} \rangle\right)^2 \right\rangle = \left\langle (\Delta x_{n,0})^2 \right\rangle - \langle \Delta x_{n,0} \rangle^2 = \left\langle \left(\sum_{j=1}^{n} \Delta x_j\right)^2 \right\rangle - \left\langle \sum_{j=1}^{n} \Delta x_j \right\rangle^2$$

$$= \left\langle \sum_{i=1}^{n} \sum_{j=1}^{n} \Delta x_i \Delta x_j \right\rangle \quad - \quad \left\{ \sum_{j=1}^{n} \langle \Delta x_j \rangle \right\}^2 \tag{5.3}$$

$$\Downarrow \qquad\qquad\qquad\qquad \Downarrow$$

$$\sum_{i=1}^{n} \left\langle (\Delta x_i)^2 \right\rangle + \sum_{i \neq j} \langle \Delta x_i \Delta x_j \rangle \qquad \sum_{i=1}^{n} \langle \Delta x_i \rangle^2 + \sum_{i \neq j} \langle \Delta x_i \rangle \langle \Delta x_j \rangle$$

Collecting up the cross terms, gives contributions of the form $\langle \Delta x_i \Delta x_j \rangle - \langle \Delta x_i \rangle \langle \Delta x_j \rangle$ which is a crucially important quantity. It connects past outcomes to current outcomes (i.e. $\Delta x_i$ at step $i$ to $\Delta x_j$ at step $j$). It will only be non-zero if there is some kind of memory (i.e. *correlation*) in the process. This is not the case for a coin toss since a coin has no memory. However, more generally there will be some kind of memory – albeit subtle – and hence these terms play a crucial role in either *increasing* or *decreasing* the variance, hence increasing or decreasing the typical walk size, and hence affecting the *b* value in the particular way discussed below. These terms are called correlation terms, which is why a system having correlations is often referred to as a system with memory.

Specifically, if the changes in the Red Queen's lead $\Delta x_i$ are *uncorrelated* (i.e. no memory) then $\langle \Delta x_i \Delta x_j \rangle = \langle \Delta x_i \rangle \langle \Delta x_j \rangle$ for $i \neq j$ and hence Equation (5.3) simplifies exactly to:

$$\sigma_{n,0}^2 = \sum_{i=1}^n \langle (\Delta x_i)^2 \rangle - \sum_{i=1}^n \langle \Delta x_i \rangle^2 = \sum_{i=1}^n \left\{ \langle (\Delta x_i)^2 \rangle - \langle \Delta x_i \rangle^2 \right\} = \sum_{i=1}^n \sigma_{i,i-1}^2 \quad (5.4)$$

Hence we have proved the well-known statistical result for *uncorrelated* variables (i.e. no memory) that the *variance of the sum is equal to the sum of the variances*. For the special case in which each variance is the same for each step (for example, for i.i.d. variables) then $\sigma_{i,i-1}^2 \equiv \sigma^2$ and we have:

$$\sigma_{i,i-n}^2 \equiv \sum_{i=1}^n \sigma_{i,i-1}^2 = n\sigma^2 \quad (5.5)$$

where $\sigma_{i,i-n}^2 = \sigma_{n,0}^2$ since the lead-changes at each step have the same variance. Taking the square root of each side of Eq. (5.5), this gives the result quoted in the main paper that the typical size (i.e. root-mean-square) of the Red Queen's walk over the *n* fatal days of the conflict (i.e. steps) increases as $n^b$ where $b = 0.5$. Using the more general notation of this Appendix, we have:

$$\sigma_{i,i-n} = n^{\frac{1}{2}} \sigma \quad (5.6)$$

So, *the typical size of the Red Queen's lead over the entire conflict to date with n fatal days, increases as the square-root of n, i.e. n to the power 0.5*. For the special case of our coin-toss walk, we have $\sigma = d$ and hence $\sigma_{i,i-n} = n^{\frac{1}{2}} d$.

In the *opposite* limit where all the changes $\Delta x_i$ are *so correlated* that they all have the same value and sign $\Delta x$ (i.e. there is memory), it then follows that

$$\sigma_{i,i-n}^2 \equiv \langle (\Delta x_{n,0} - \langle \Delta x_{n,0} \rangle)^2 \rangle = \langle (\Delta x_{n,0})^2 \rangle - \langle \Delta x_{n,0} \rangle^2 = \left\langle \left( \sum_{j=1}^n \Delta x_j \right)^2 \right\rangle - \left\langle \sum_{j=1}^n \Delta x_j \right\rangle^2$$

$$= \langle (n \Delta x)^2 \rangle - \langle n \Delta x \rangle^2 = n^2 \left( \langle (\Delta x)^2 \rangle - \langle \Delta x \rangle^2 \right) \quad (5.7)$$

$$= n^2 \sigma^2$$

and hence the typical size of the Red Queen's lead over the entire conflict to date with *n* fatal days, now increases as

$$\sigma_{i,i-n} = n \sigma \quad (5.8)$$

as stated in the main paper. In other words, the typical size of the Red Queen's walk over *n* fatal days (i.e. steps) increases as $n^b$ where $b = 1$. This makes sense: think of walking purposely in a straight line at constant velocity. The distance moved is now proportional to the number of timesteps *n*, i.e. $n^{b=1}$ as opposed to $n^{b=0.5}$. In the more general case of some *limited* but *non-zero* level of positive correlation (i.e. some memory), the corresponding expression for the typical size of *R* (i.e. root-mean-square) will therefore lie between the uncorrelated case of $n^{b=0.5}$ and the perfectly correlated case of $n^{b=1}$. By contrast, if the changes $\Delta x_i$ are *anti-correlated* (i.e. their correlation is negative), then the dependence will be more like $n^{b=0}$ as stated in the main paper.

Hence in general (i.e. in the presence of correlations) the Red Queen's walk will have the property that the typical size of the walk after $n$ time-steps increases as $n^b$, where $b$ takes a value which is typically in the range 0 to 1, but may be higher if the feedback is positive (i.e. Red Queen is not only walking in straight line but is also accelerating as she walks). This confirms that $b = 0.5$ acts as a crude midpoint (see green dashed line in Figs. 1(c) and 3 of main paper). We note that the terminology employed in polymer physics, is that a 'persistent walk' corresponds to $b > 0.5$ while an 'anti-persistent walk' corresponds to $0 \leq b < 0.5$.

## 5.3 Multi dimensional Red Queen walk

Although this section is likely to be of more interest to physical scientists, in particular statistical physicists including polymer physicists, we note that the present analysis of a stochastic walk in one dimension can be generalized to higher dimensions – where the individual axes in this higher dimensional space represent different types of strategic advantage. For completeness, we now present this discussion though we stress that it is not needed for the general reader's understanding of the main paper. Following the $n$'th successful event, the relative separation between the Red Queen's and Blue King's strategies, tactics and/or technological advantage, is now a vector $\vec{R}(n)$ in some $D$-dimensional advantage space. $\vec{R}(n)$ changes by one step within this $D$-dimensional space, i.e. the stochastically varying vector $\vec{R}(n)$ moves in a general direction at each event step $n$. Borrowing from Physics (*27*), the typical size of $|\vec{R}(n)|$ and hence the insurgents' relative advantage after $n$ successful tasks, is given by the root-mean square size of the $D$-dimensional walk, i.e. $|\vec{R}(n)|_{rms} \propto n^b$ with $b > 0$. With effective military counter-adaptation, $b \approx 0$ hence $|\vec{R}(n)|_{rms} \sim 0$. If by contrast the military never counter-adapt successfully, then the insurgents benefit from all fatal days and $b = 1$ (or $b > 1$ if insurgent progress is superlinear as a result of positive feedback). In the intermediate case, the military's counter-adaptation success would follow $|\vec{R}(n)|_{rms} \propto n^{1/2}$ which means $b = 0.5$ (*27*).

If the resulting walk never returns to the same point, then $b = 3/(2 + D)$ (*27*), implying that $b = 1$ if there is $D = 1$ dominant strategic dimension. $D = 2$ gives $b = 0.75$ etc. Making the reasonable assumption that the current relative advantage $|\vec{R}(n)|_{rms}$ is proportional to the current rate of fatal days, then the typical time interval between attacks will be proportional to $\left[|\vec{R}(n)|_{rms}\right]^{-1} \propto n^{-b}$. Since the first time interval is $\tau_1$, then the theoretical time interval $\tau_n = \tau_1 n^{-b}$ for $b \geq 0$, with a typical value around $b = 0.5$.

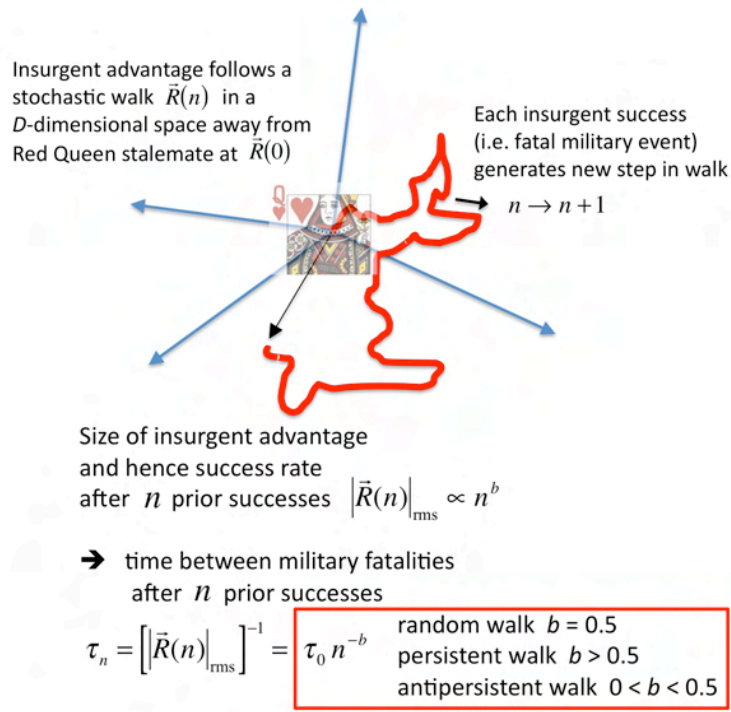

Figure caption: Our dynamic Red Queen model for insurgent attack dynamics within each province. Insurgent advantage follows a stochastic walk in *D*-dimensional space. Using known results from Physics (*27*) exact results can be obtained for *b* under different conditions (see text).